\documentclass[twocolumn]{aastex631}
\accepted{December 7, 2022}

\submitjournal{ApJ}

\shorttitle{From Centaurs to Jupiter-family Comets}
\shortauthors{Guilbert-Lepoutre et al.}

\begin{document}
\title{The gateway from Centaurs to Jupiter-family Comets: thermal and dynamical evolution}

\correspondingauthor{Aurélie Guilbert-Lepoutre}
\email{aurelie.guilbert-lepoutre@univ-lyon1.fr}

\author[0000-0003-2354-0766]{Aur\'{e}lie Guilbert-Lepoutre}
\author[0000-0002-1611-0381]{Anastasios Gkotsinas}
\affiliation{Laboratoire de G\'{e}ologie de Lyon: Terre, Plan\`{e}tes, Environnnement, CNRS, UCBL, ENSL, F-69622, Villeurbanne, France} 

\author[0000-0001-8974-0758]{Sean N. Raymond}
\affiliation{Laboratoire d'Astrophysique de Bordeaux, Univ. Bordeaux, CNRS, F-33615 Pessac, France}

\author[0000-0002-4547-4301]{David Nesvorny}
\affiliation{Department of Space Studies, Southwest Research Institute, 1050 Walnut Street, Suite 300, Boulder, CO 80302, USA}

\begin{abstract}

It was recently proposed that there exists a ``gateway'' in the orbital parameter space through which Centaurs transition to Jupiter-family Comets (JFCs).  
Further studies have implied that the majority of objects that eventually evolve into JFCs should leave the Centaur population through this gateway. 
This may be naively interpreted as gateway Centaurs being pristine progenitors of JFCs. This is the point we want to address in this work.
We show that the opposite is true: gateway Centaurs are, on average, {\em more} thermally processed than the rest of the population of Centaurs crossing Jupiter's orbit. Using a dynamically-validated JFC population, we find that only $\sim 20\%$ of Centaurs pass through the gateway {\em prior} to becoming JFCs, in accordance with previous studies.
We show that more than half of JFC dynamical clones entering the gateway for the first time have already been JFCs -- they simply avoided the gateway on their first pass into the inner solar system. 
By coupling a thermal evolution model to the orbital evolution of JFC dynamical clones, we find a higher than 50\% chance that the layer currently contributing to the observed activity of gateway objects has been physically and chemically altered, due to previously sustained thermal processing. We further illustrate this effect by examining dynamical clones that match the present-day orbits of 29P/Schwassmann-Wachmann~1,  P/2019~LD2~(ATLAS), and P/2008~CL94~(Lemmon).

\end{abstract}


\section{Introduction} \label{sec:intro}
\begin{figure}[!h]
    \centering
    \includegraphics[width=\columnwidth]{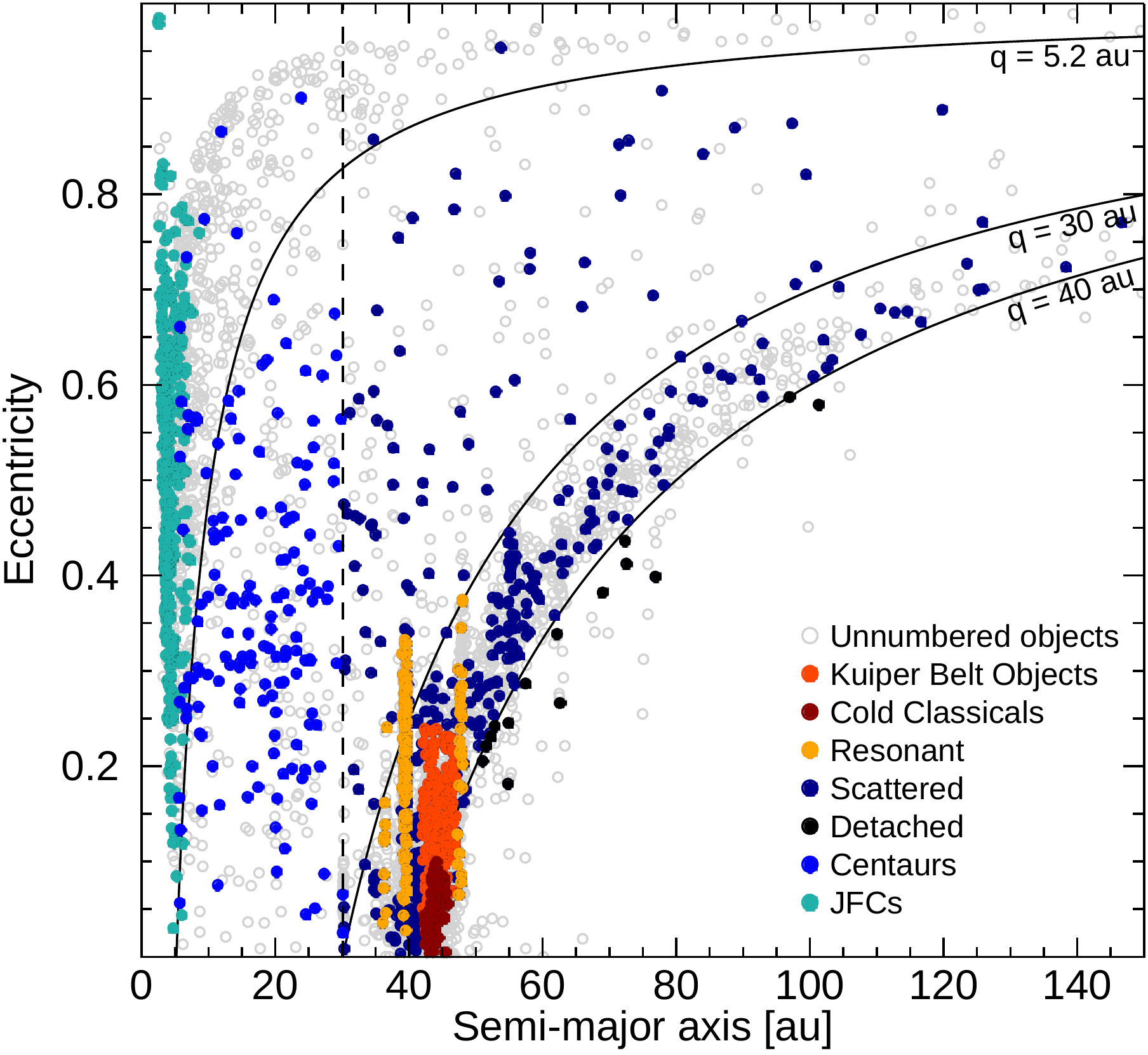}
    \includegraphics[width=\columnwidth]{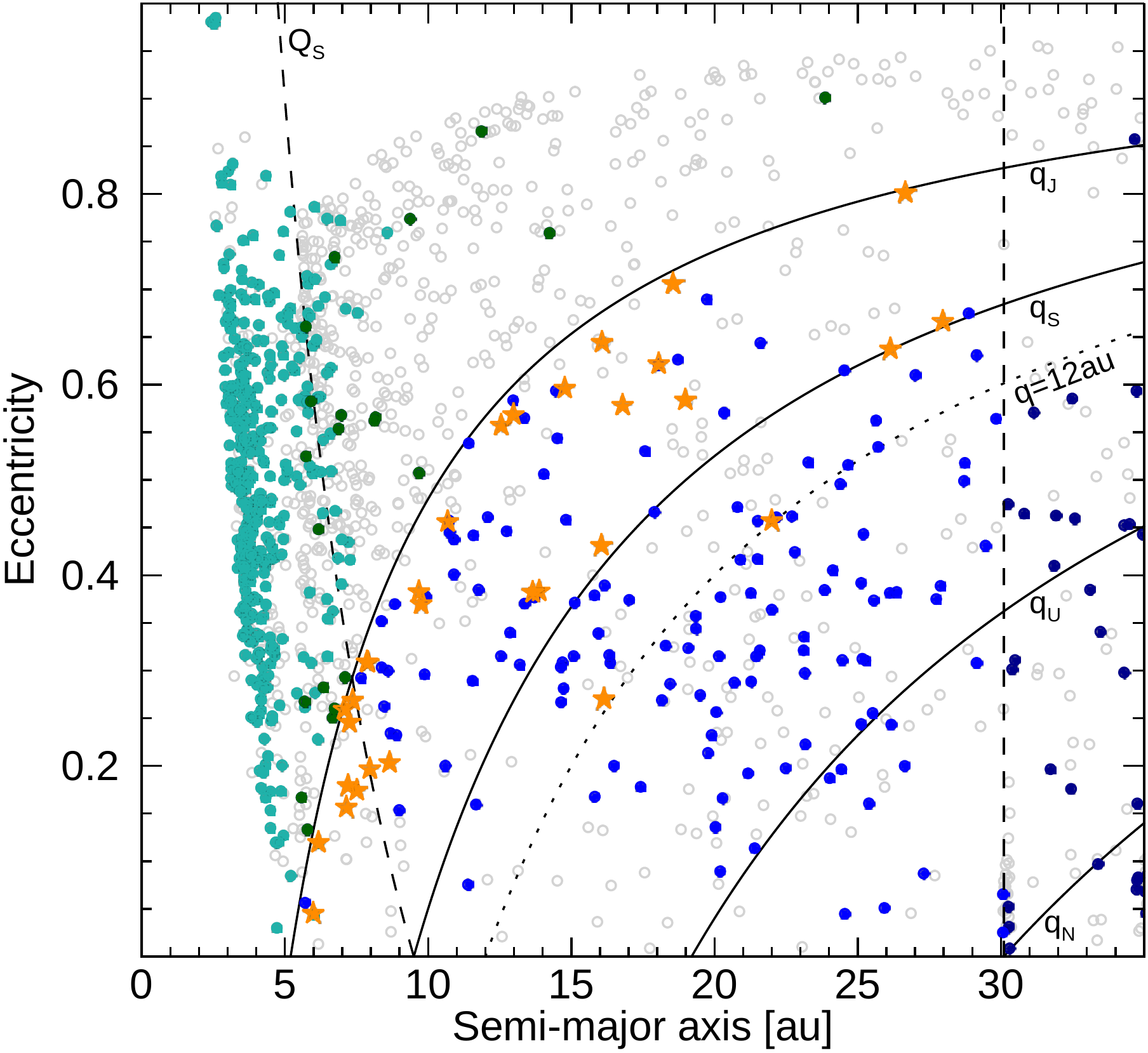}

    \caption{Distribution of icy objects in the solar system, from the trans-neptunian to the JFC populations, in the semi-major axis (in au) vs. eccentricity plane. Neptune's orbit is marked by a vertical dashed line. Other lines of interest are displayed: on the top panel, the locus of perihelion distances at 5.2~au, 30~au, 40~au; on the bottom panel, the locus of perihelion distances at the giant planets and 12~au, and the locus of aphelion distance at Saturn. Orbital elements come from the JPL Solar System Dynamics database (\textit{https://ssd.jpl.nasa.gov}). Active Centaurs are marked with orange stars on the bottom panel.}
    \label{populations}
\end{figure}

Jupiter-family Comets (JFCs) are continuously replenished from their outer solar system reservoirs, the Kuiper Belt and the scattered disk \citep[][see distributions in Fig.\ref{populations}]{Fernandez1980, Duncan1988}. Before JFCs enter the inner solar system, where they are typically observed on short-period orbits with perihelion distances close to the Sun, they spend a significant amount of time as Centaurs \citep{Levison1997, Tiscareno2003}. This dynamical cascade between populations, and the individual orbital tracks that these icy objects follow, can entail extensive modifications of their internal structure and composition \citep[e.g.][and references therein]{Gkotsinas2022}. In this context, the transient population of Centaurs is a key target for understanding progenitors of JFCs. 

Recently, \citet{Sarid2019} reported that the transition from the Centaur to the JFC region involves a passage through a restricted area in the orbital elements space, described by orbits with a perihelion distance $q>$~5.4~au and an aphelion distance $Q<$~7.8~au, which translates into orbits with semi-major axis 5.2~$<a<$~7.8~au and an eccentricity $e<$~0.2. 
This ``gateway'' region has a heliocentric distance range that coincides with that where cometary nuclei are observed to be increasingly active within the giant planet region (see active Centaurs displayed with orange stars in Fig.\ref{populations}). Their dynamical models suggest that the majority of objects which eventually become JFCs transition from the Centaur population through this gateway. Subsequent work have emphasized the importance of studying the activity pattern of so-called gateway Centaurs, as they will likely transition to JFCs in relatively short timescales, becoming ideal targets to investigate how dynamical and thermal evolution alters comet nuclei before becoming JFCs \citep[e.g.][]{Steckloff2020, Kareta2021}.

In this work, we examine the thermal processing of objects transitioning from the Centaur to the JFC region, with an emphasis on the gateway region. We use a sample of simulated JFCs (hereafter dynamical clones), successfully reproducing the current orbital distribution of JFCs, taken from a N-body simulation tracking the orbital evolution of the giant planets, and a large number of small bodies of the outer planetesimal disk \citep{Nesvorny2017}. We apply a thermal evolution model \citep{Guilbert2011} to the resulting orbital evolution tracks, in order to constrain the internal thermal structure -- a method developed by \citet{Gkotsinas2022}. Our goal is to assess the significance of this region for the physical properties of active Centaurs, as they evolve from the outer solar system to Jupiter-crossing orbits.

\section{Coupled thermal and dynamical evolution study}

\subsection{Dynamical clones}

We consider in this study a sample of JFC dynamical clones generated from simulations performed by \citet{Nesvorny2017}. 
The goal of their work was: a) to model the formation of cometary reservoirs early in the solar system history ; b) follow their evolution up to the present time ; c)  assess how current observations of well characterized JFCs could be used to constrain the orbital structure of the transneptunian region. In order to do so, they performed end-to-end simulations, forming cometary reservoirs and letting them evolve for 4.5~Gyr. 
These simulations rely on a dynamical framework for the early evolution of the solar system, including the planetary migrations and instabilities which lead to the solar system as we know it now. They used the model described by \citet{nesvorny2012}, for which self-consistent simulations were performed and tested against various constraints from small body populations \citep[e.g. asteroids, Kuiper Belt, Jupiter Trojans, regular and irregular moons of the giant planets,][]{nesvorny2018}.

\citet{Nesvorny2017} calibrated their model by confronting the characteristics of their JFC dynamical clones (e.g. number, orbital element distributions) to observed comets. The observable JFCs, with a perihelion distance below 2.5~au, amount to 350 to 380 objects currently known and well-enough characterized to be used to perform such calibration \citep[see also][]{Seligman2021}.
For each resulting dynamical clone, they record the dynamical pathway from the time it leaves the reservoir, until it is ejected out of the solar system. These trajectories can thus be used to study the dynamical and physical evolution of JFCs in a statistically significant manner. We note that non-gravitational forces were ignored. The timestep for the simulations is 0.5~yr, however the trajectories themselves are recorded every 100~yr from the first time the clones reach 30~au on their way inward, out of the outer solar system's reservoirs.

\subsection{Coupled thermal and orbital evolution}

For this study, we consider a total of 350 JFC dynamical clones, all of which have a perihelion distance within 2.5~au at some point in their lifetime. 
Coupling the thermal evolution of these clones to their orbital evolution requires making a number of assumptions to constrain their effective long-term thermal processing, which we describe below.

\paragraph{Heat equation} ~
We use a 1D thermal evolution model derived from \citet{Guilbert2011}, which solves the heat diffusion equation:
\begin{equation}\label{eq:heateq}
    \rho_{bulk}c~ \frac{\partial T}{\partial t} + div(-\kappa ~\overrightarrow{grad}~ T) = \mathcal{S},
\end{equation}
with $\rho_{bulk}$ (kg m$^{-3}$) the clone's bulk density, $c$ (J kg$^{-1}$ K$^{-1}$) the material's heat capacity, $T$ (K) the temperature, $\kappa$ (W m$^{-1}$ K$^{-1}$) the material's effective thermal conductivity, and $\mathcal{S}$ the heat sources and sinks.
First, we exclude phase transitions, which we cannot track properly because they occur on timescales much smaller than the dynamical timestep \citep{Gkotsinas2022}. Moreover, our thermal evolution model would require a prohibitive calculation time to solve time-dependent equations of heat transfer and gas flow in a porous medium, while accounting for multiple phase transitions, during the millions of years achieved by dynamical simulations. In Equation~\ref{eq:heateq}, this leads to $\mathcal{S}=0$.

\paragraph{Physical properties} ~ 
Each dynamical clone is considered as a sphere with a 5~km-radius. This parameter has no influence on our results, as the only source of heating we consider is insolation of the surface. The size of clones ($R$, in m) would affect the conduction timescale $\tau = R^2 \rho_{bulk} c / \kappa$ (s), which informs on the time required to heat an object down to the core, and which is much longer than the orbital evolution timescale (for km-sized objects). Essentially, we are interested in the processing of the subsurface layer which contributes to any activity observed today, i.e. a few hundred meters at most. All physical characteristics are assumed to remain constant through the dynamical evolution. For each parameter, we select a reference value extensively used in the literature \citep[e.g.][]{Prialnik2004, Huebner2006}. The most critical of those is the thermal conductivity $\kappa$ (W~m$^{-1}$~K$^{-1}$), which defines the fraction of heat diffusing toward the interior \citep[see][for the influence on the long-term processing of JFCs]{Gkotsinas2022}. Different values of the thermal conductivity ultimately result in different depths at which heat waves can penetrate below the surface, e.g. a lower conductivity induces the processing of  a shallower subsurface layer. This effect does not modify the heating patterns that we describe though, nor the general conclusion as subsequent activity generated from the subsurface layers is also scaled with the thermal conductivity. Therefore, in the following, we only show results obtained with a thermal conductivity of 5$\times$10$^{-3}$~W~m$^{-1}$~K$^{-1}$, a realistic value in agreement with laboratory experiments on cometary material \citep[i.e. 0.002~$<\kappa<$~0.02~W~m$^{-1}$~K$^{-1}$,][]{Krause2011}.

\paragraph{Energy balance at the surface} ~
Further simplifications are adopted regarding the calculation of the energy balance at the surface over the 100~yr dynamical time step. An averaged energy flux is computed at every time step, based on analytical solutions for the time averaged energy flux received by objects on eccentric orbits \citep[][]{Williams_2002, Mendez_2017}. This averaged energy flux defines an averaged orbital distance computed as $a_{eq} = a(1-e^2)$ (au), which is used in the surface boundary condition of Equation \ref{eq:heateq}:
\begin{equation}
     \frac{(1-\mathcal{A}) L_\odot}{4\pi a_{eq}^2} = \varepsilon \sigma_{SB} T^4 + \kappa \frac{\partial T}{\partial r} \label{eq:boundary}
\end{equation}
with $\mathcal{A}$ the Bond's albedo, $L_\odot$ the solar constant, $\varepsilon$ the emissivity, $\sigma_{SB}$ the Stefan-Boltzmann constant. Diurnal and seasonal variations are not considered in our thermal simulations, as they are simply not resolved in such long-term dynamical simulations. Additional limitations arise when it comes to sharp orbital changes taking place mainly in the inner parts of the solar system, close to Jupiter and Saturn \citep{Seligman2021}: these are not resolved by the 100~yr time step in the dynamical simulations outputs. This implies that some short-scale heating episodes will go unnoticed in our thermal simulations. However, internal heating on such short timescales is likely limited to a shallow subsurface layer. Indeed, an intense but quick passage close to the Sun has a limited effect on a comet's interior than a lengthier exposure to an averaged lesser amount of energy received at the surface. Our averaging strategy over a 100~yr time step thus mitigates the effects of both short and long exposures to insolation.

\section{Transition of clones between Centaurs and JFCs}

\subsection{Definition of populations}
To investigate the transition of icy objects between the Centaur and JFC populations, we first need to define the contours of these populations. To do so, we put a label on these bodies, based on cuts and thresholds in the distribution of their orbital elements, despite the clear continuity between populations (see Fig.\ref{populations} for instance).
In other words, these definitions do not inform, or alter, the nature of these objects. Many definitions can be found in the literature, as recently reviewed by \citet{Seligman2021}. 
In this study, we based our definitions for Centaurs and JFCs on the definition of the Gateway as given in \cite{Sarid2019}. 
We remind that it is introduced as orbits which do not cross the orbit of Jupiter, i.e. with a perihelion distance $q>$~5.4~au. Objects in the Gateway should also be well separated from the orbit of Saturn, i.e. they should have an aphelion distance $Q<$~7.8~au.
We thus define the Centaur population as having 5.4~$<q<$~30.1~au and 5.4~$<a<$~30.1~au, which is relatively similar to the comprehensive definition of \citet{Jewitt2009}. Consequently, Centaurs which are not in the gateway have $q>$~5.4~au, and $Q>$~7.8~au, and JFCs are objects with $q<$~5.4~au and $Q<$~7.8~au. With these definitions, a number of objects (or rather, orbits explored by clones) do not find any ``host'' population, because their orbital elements do not satisfy the thresholds to receive the corresponding label. Indeed, a number of clones transition to the JFC population from regions in the orbital space that do not fit the above cuts, due mainly to a $e>$~0.3 eccentricity, allowing clones with a large semi-major axis to reach Jupiter-crossing orbits.
Hence, to achieve a complete description of the distribution, we define a population of Jupiter-crossers, with $q<$~5.4~au, and 7.8~$<Q<$~14.5~au. 
This 14.5~au threshold is based on the consideration that $a$ should be smaller than the semi-major axis of Saturn (so that the orbital evolution is dominated by interactions with Jupiter).
We note that this category is relevant in particular for objects which never go through the gateway region during their lifetime.

\subsection{To go or not go through the gateway}

\begin{figure}[!h]
    \centering
    \includegraphics[width=\columnwidth]{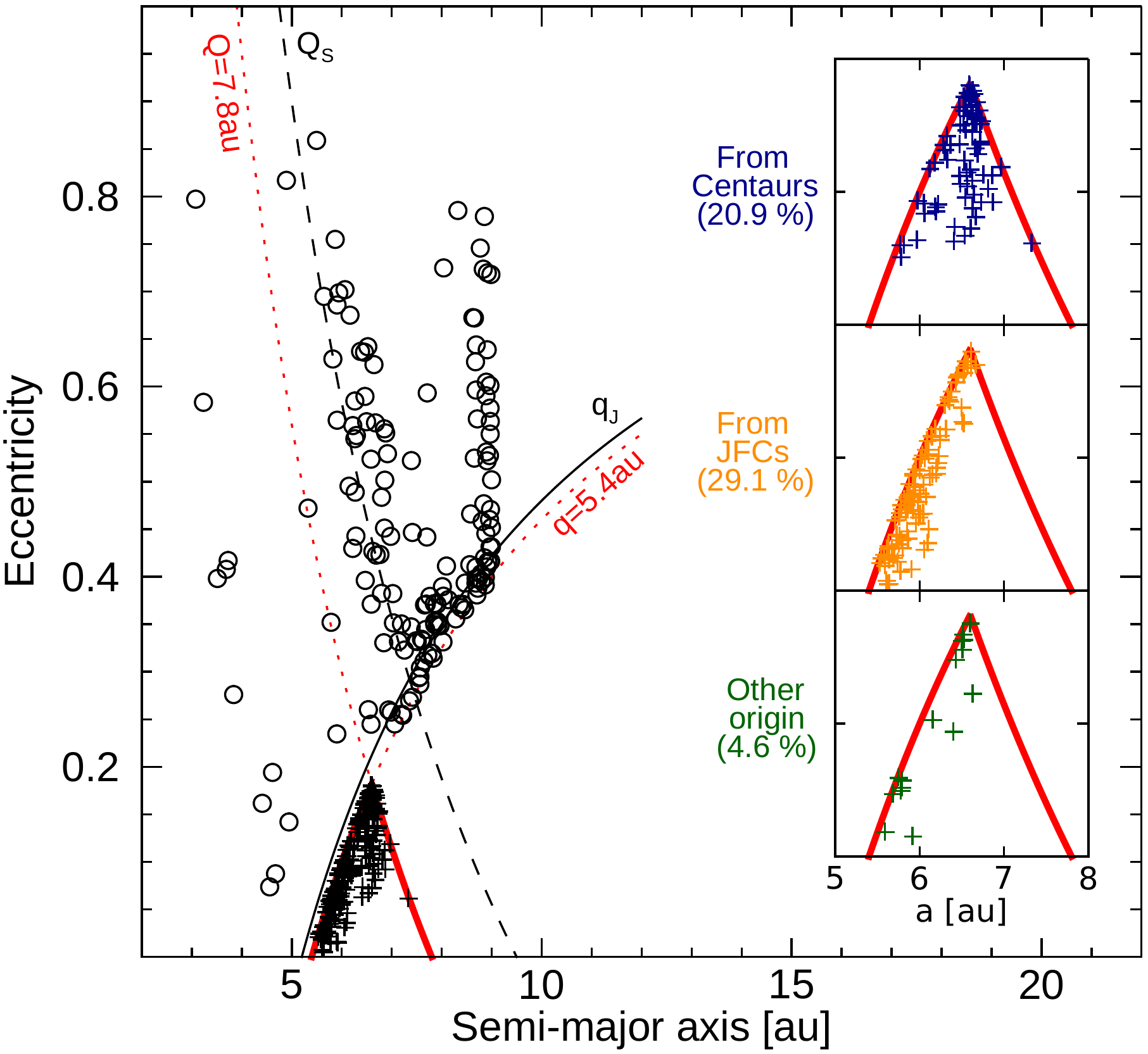}
    \caption{Distribution of dynamical clones in the semi-major axis vs. eccentricity plane on the first time they enter the gateway or cross Jupiter's orbit. Crosses correspond to clones (191 objects, 54.6\% of the population) entering the gateway for the first time. Distributions as a function of their origin are given on the right panels (Centaurs in blue, JFCs in orange, or Jupiter-crossers in  green). Circles correspond to the 159 clones (45.4\%) which never go through the gateway: their distribution is given on the first time they become Jupiter-crossers before becoming JFCs.}
    \label{gateway}
\end{figure}

With these thresholds in mind, we investigate the population of dynamical clones from \citet{Nesvorny2017} as they transition from Centaurs to JFCs. Among the 350 clones, we find that 191 objects reach the gateway region at least once in their lifetime (54.6\%). Of those, 73 were Centaurs prior to entering the gateway (i.e. 20.9\% of the overall clone population), while 102 objects (29.1\%) were previously JFCs. In other words, these clones had already transitioned from Centaurs to JFCs without going through the gateway, which they entered then later during their lifetime. The remainder 16 clones (4.6\%) entered the gateway from Jupiter-crossing orbits. The distribution of these objects is given in Fig.\ref{gateway}. 

Overall, we find that strictly speaking, our population has only 20.9\% of Centaurs which actually go through the gateway \textbf{prior} to becoming JFCs for the first time.
Since 159 clones (45.4\%) never go through the gateway at any point of their lifetime, we find that most Centaurs (79.1\%) make their first transition to the JFC population outside of the gateway region.
As reported by \citet{Sarid2019}, we find that an object -- of those reaching the gateway region -- can enter and exit the gateway more than once during its lifetime: the mean number of entrance is 7 to 8 times, the median is at 4 entrances though.

\subsection{Thermal processing of clones in the gateway}

In the same way as \citet{Gkotsinas2022}, we assess the thermal processing of our dynamical clones in a statistical manner, by tracking the depth of three isotherms representative of key phase transitions: a) 25~K, for the sublimation of hypervolatile species such as CO ; b) 80~K, for the sublimation of moderately volatile species such as CO$_2$ ; c) 110~K, for the crystallization of amorphous water ice. 
For clones which go through the gateway at some point of their lifetime, we record the depth of those isotherms on the first time they enter this region: their corresponding orbital elements are shown in Fig.\ref{gateway}.
For clones which never go through the gateway, we record these depths on the first time they cross the orbit of Jupiter, i.e. the first time their orbit satisfies $q<$~5.4~au and 7.8~$<Q<$~14.5~au. The distribution of orbital elements of these clones in the semi-major axis vs. eccentricity plane is given in Fig.\ref{gateway}.
\begin{figure*}[!ht]
    \centering
    \includegraphics[width=\textwidth]{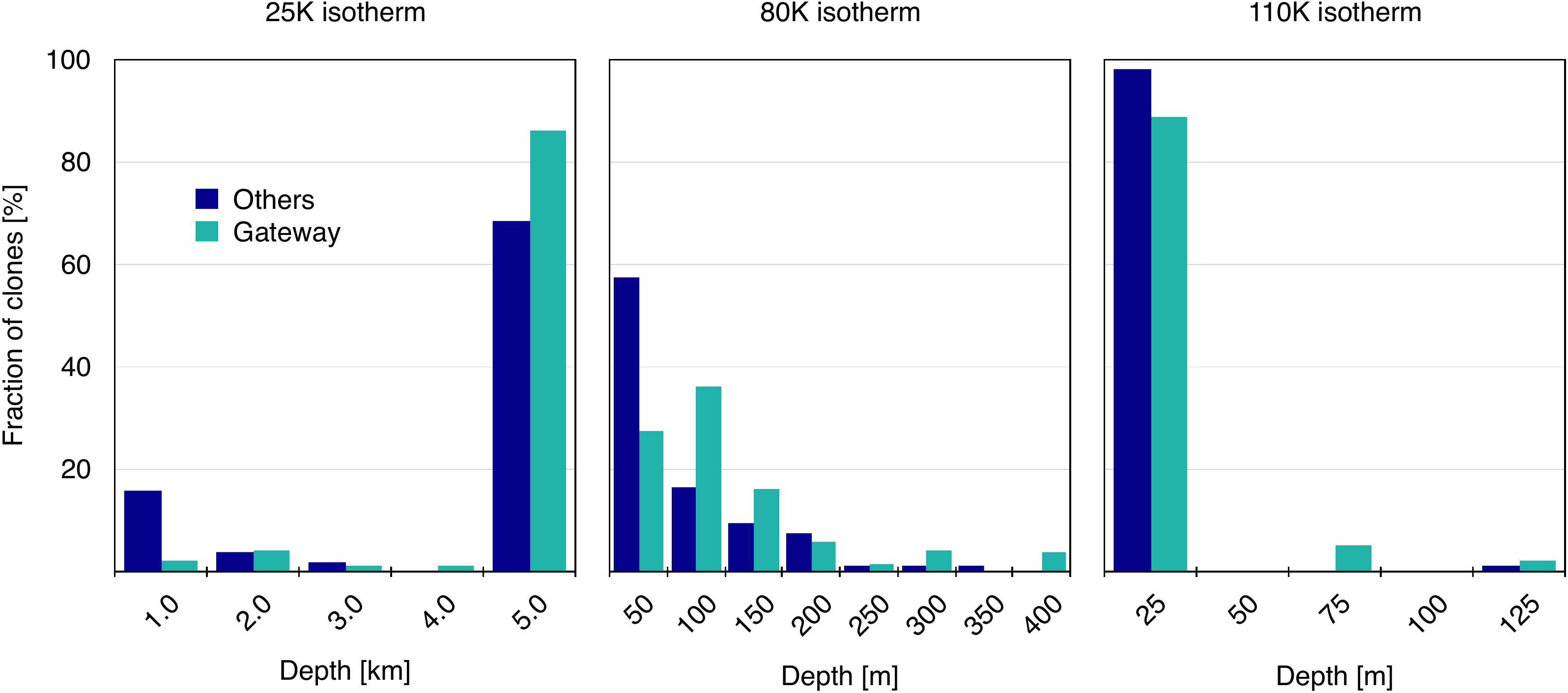}
    \caption{Temperature distributions for clones when they enter the gateway for the first time (teal), and clones when they first cross Jupiter prior to reaching JFC orbits (dark blue).}
    \label{depth}
\end{figure*} 
We show the corresponding internal temperature distributions in Fig.\ref{depth}.
We find some differences in the thermal processing of the two sub-populations, i.e. clones passing through the gateway vs. clones reaching JFC orbits without ever entering the gateway in their lifetime. 

Indeed, objects are statistically more processed on their first entrance in the gateway than the rest of Centaurs, when they transition to JFC orbits outside of the gateway.
We applied a Mann-Whitney U non-parametric test for each isotherm: this test can be used for non-normal distributions of populations with different variances, to test the null hypothesis that two samples come from the same population.
Each test confirms that the two groups are statistically different (null hypothesis rejected with a p-value of 10$^{-6}$ for 25~K, and 10$^{-8}$ for 80 and 110~K). 
This is mainly due to the fact that more than half of these objects (102 clones or 53.4\% of objects going through the gateway) have already been close to the Sun on JFC orbits, prior to reaching the gateway. In contrast, the ``no-gateway'' objects are considered on their first time of crossing the orbit of Jupiter, prior to becoming JFCs in this comparison. The gateway clones are thus more processed on average than those of Centaur origin: their internal structure and composition is affected by the cumulative effect of experiencing higher temperatures on JFC orbits, and having spent more time (on average) close to the Sun.
This effect was accounted for by \citet{Sarid2019} through a fading activity law. 

If most dynamical clones are heated above 25~K down to the core, the ``no-gateway'' population has the largest fraction of objects able to maintain some hypervolatiles (as pure condensates) within the 1~km-subsurface layer (see Fig.\ref{depth} and Table \ref{tab:stat}). 
Similarly, the 80~K isotherm is located on average $\sim$60~m below the surface for the ``no-gateway'' clones (median around 10~m), while the gateway population is heated above that temperature for more than 100~m on average (median at 60~m). The crystallization front (represented by the 110~K isotherm) remains close to the surface for both sub-populations, although more objects in the gateway experience crystallization below 50-100~m than the ``no-gateway'' objects. 

\begin{table}[h!]
    \centering
    \begin{tabular}{lcc}
     \hline
     \multicolumn{3}{c}{Gateway Centaurs (191 objects)}\\
     \hline
     Isotherm      & Average depth (m) & Median depth (m) \\
     25 K  & 4760 & 5000 \\     
     80 K  & 110  & 61   \\
     110 K & 20   & 4.4  \\
     \hline      
     \multicolumn{3}{c}{``No-gateway'' Centaurs (159 objects)}\\
     \hline
     Isotherm      & Average depth (m) & Median depth (m) \\
     25 K  & 3810 & 5000\\     
     80 K  & 62   & 11  \\
     110 K & 4.2  & 0.25\\         
    \end{tabular}
    \caption{Depth of the 25K, 80K and 110K isotherms for the gateway and ``no-gateway'' Centaurs. The averages and medians are computed from the temperature distributions shown in Fig.\ref{depth}}
    \label{tab:stat}
\end{table}

\section{Implication for individual objects}

\subsection{Context of active Centaurs}
As of today, the origin of Centaurs' activity has not been definitively identified, and different processes may be involved for different individual objects \citep[e.g.][]{Prialnik1995, Capria2000, DeSanctis2000}. Crystallization of amorphous water ice appears as a phase transition of choice to trigger activity, given the physical and orbital properties of active Centaurs \citep[][see also Fig.\ref{populations}]{Jewitt2009}:
indeed, currently known active Centaurs are too cold for water ice to sublimate, while the sublimation of other species such as CO or CO$_2$ would imply that activity should be observed even further out in the giant planet region.
Overall, \citet{Guilbert2012} suggested that the activity of Centaurs seems tightly linked to their orbit: as amorphous water ice crystallization progresses inward below the surface, sustained cometary activity fades with time. A change in surface energy balance (e.g. due to a drop in perihelion distance) is thus required to trigger a subsequent new spurt of activity.
\citet{Davidsson2021} argued that the sublimation and segregation of CO$_2$ may additionally play some role to explain the level of activity observed in the 10-12~au region. 

\citet{Fernandez2018} studied the dynamical evolution of both active and inactive Centaurs. They found that active Centaurs are prone to drastic drops in their perihelion distances, with a timescale of 10$^2$ to 10$^3$~yrs. Thermal results from \citet{Guilbert2012} are consistent with these timescales, as they suggested that a change in orbital elements might be required to trigger phase transitions, ensuing the adjustment to new thermal conditions. We note that searches for activity amongst recently discovered Centaurs have failed \citep[e.g.][]{Cabral2019, Li2020, Lilly2021}, however, targeted objects are found on relatively stable orbits beyond Saturn, where no significant activity would be expected from the aforementioned processes.

With these considerations in mind, we find it relevant to study the coupled thermal and dynamical evolution of individual objects \textbf{before} they evolve to the orbit on which they are currently observed, to inform on their possible past history and activity.
We put an emphasis on the 80 and 110~K isotherms, representative of the CO$_2$ sublimation and amorphous to crystalline phase transitions, respectively.

\subsection{Selecting clones for Centaurs of interest}
The study of gateway Centaurs has focused so far on two specific objects, 29P/Schwassmann-Wachmann~1 and P/2019~LD2~(ATLAS) \citep[hereafter 29P and LD2 respectively,][]{Sarid2019, Steckloff2020, Kareta2021, Hsieh2021, Seligman2021}. In light of results presented above, we provide some insight on the coupled thermal and dynamical evolution of clones of these two bodies. 
We have added P/2008~CL94~(Lemmon) in this work (hereafter CL94), since it is an active Centaur currently located in the gateway \citep{Kulyk2016}.  
Clones of a specific object can be selected from the whole population by defining ``boxes'' around the values of currently observed semi-major axis $a$, eccentricity $e$ and inclination $i$. 
For each orbital element, we define an acceptable range of tolerance, where our dynamical clones can fall: the larger the acceptable range, the larger the number of clones that will satisfy the conditions at some point of their orbital evolution. We typically allow $\pm$0.05~au for the semi-major axis, $\pm$0.05 for the eccentricity, and $\pm$1$^o$ for the inclination. For each of the three objects mentioned above, we thus define the following ``boxes'':
\begin{itemize}
    \item[29P] \begin{itemize}
                     \item[] 5.95~$<a<$~6.05~au
                     \item[] 0.01~$<e<$~0.09
                     \item[] 8.73~$<i<$~10.73$^o$
                \end{itemize} 
    \item[LD2] \begin{itemize}
                     \item[] 5.24~$<a<$~5.34~au
                     \item[] 0.08~$<e<$~0.18
                     \item[] 10.56~$<i<$~12.56$^o$
                \end{itemize}   
    \item[CL94] \begin{itemize}
                     \item[] 6.10~$<a<$~6.20~au
                     \item[] 0.07~$<e<$~0.17
                     \item[] 7.3~$<i<$~9.3$^o$
                \end{itemize}            
\end{itemize}

We find 10 clones which at some point of their lifetime had orbital elements similar to the observed 29P and LD2. For CL94, we find 21 clones whose orbital elements satisfy the conditions at some point of their lifetime. In the following, we only consider the 
orbital history of each object before it evolves into the relevant box, as a proxy for the past orbital history of these Centaurs.
As a sanity check, we have searched for clones of each known active Centaur (i.e. 31 additional Centaurs, as displayed in Fig.\ref{populations} bottom panel), with the same accepted range on each orbital element. For each active Centaur, we find a number of clones that satisfy our conditions: on average 18 clones per object, ranging from 2 clones for C/2012~Q1~(Kowalski) to 50 clones for P/2010~C1~(Scotti) (median at 14). We thus conclude that there is nothing particular with the orbital boxes defined for 29P, LD2 or CL94, that would have led to a significantly different number of clones than the rest of active objects in the giant planet region.

\subsection{Orbital considerations}

\begin{figure}[!h]
   \centering
   \includegraphics[width=\columnwidth]{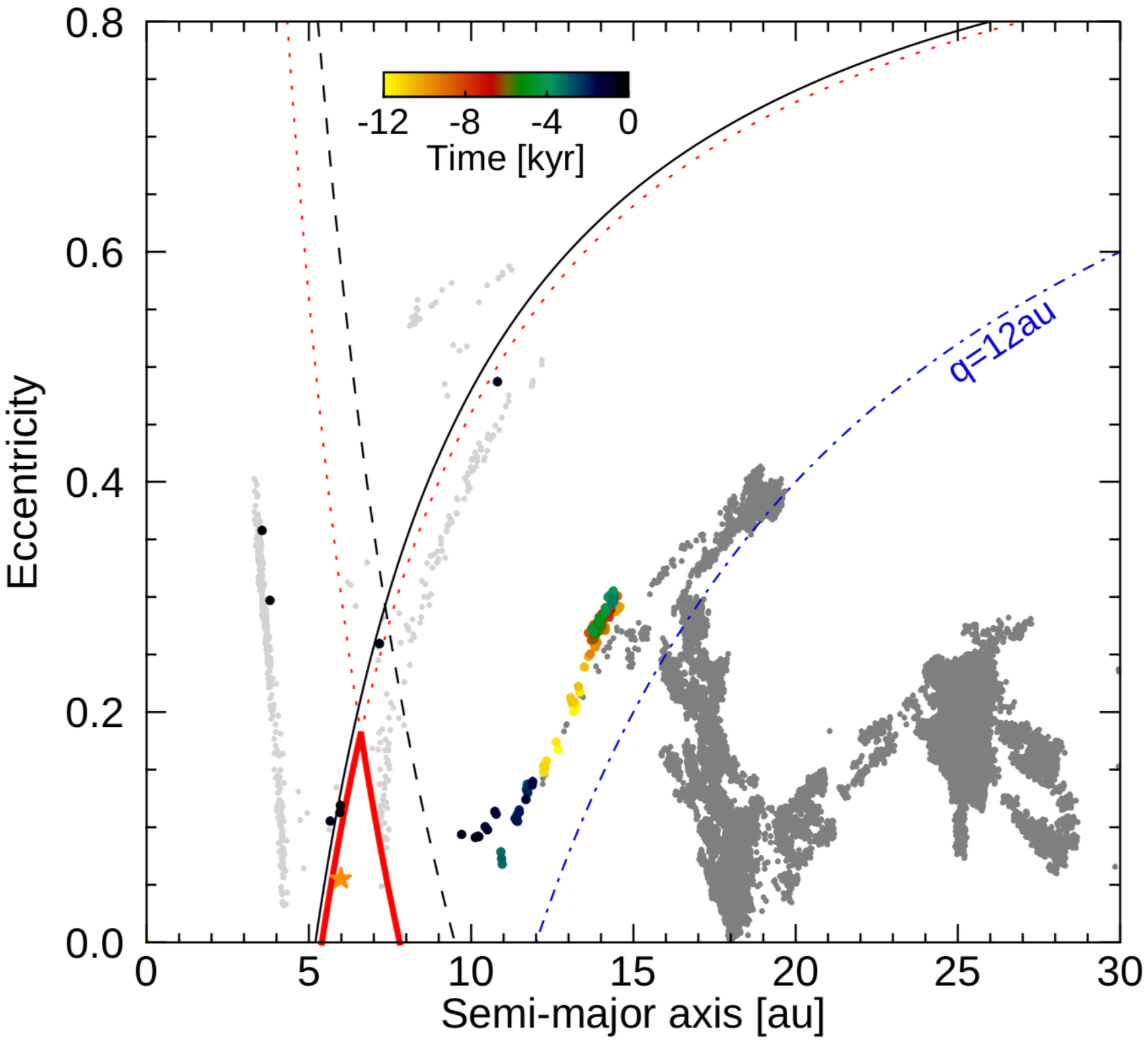}
   \includegraphics[width=\columnwidth]{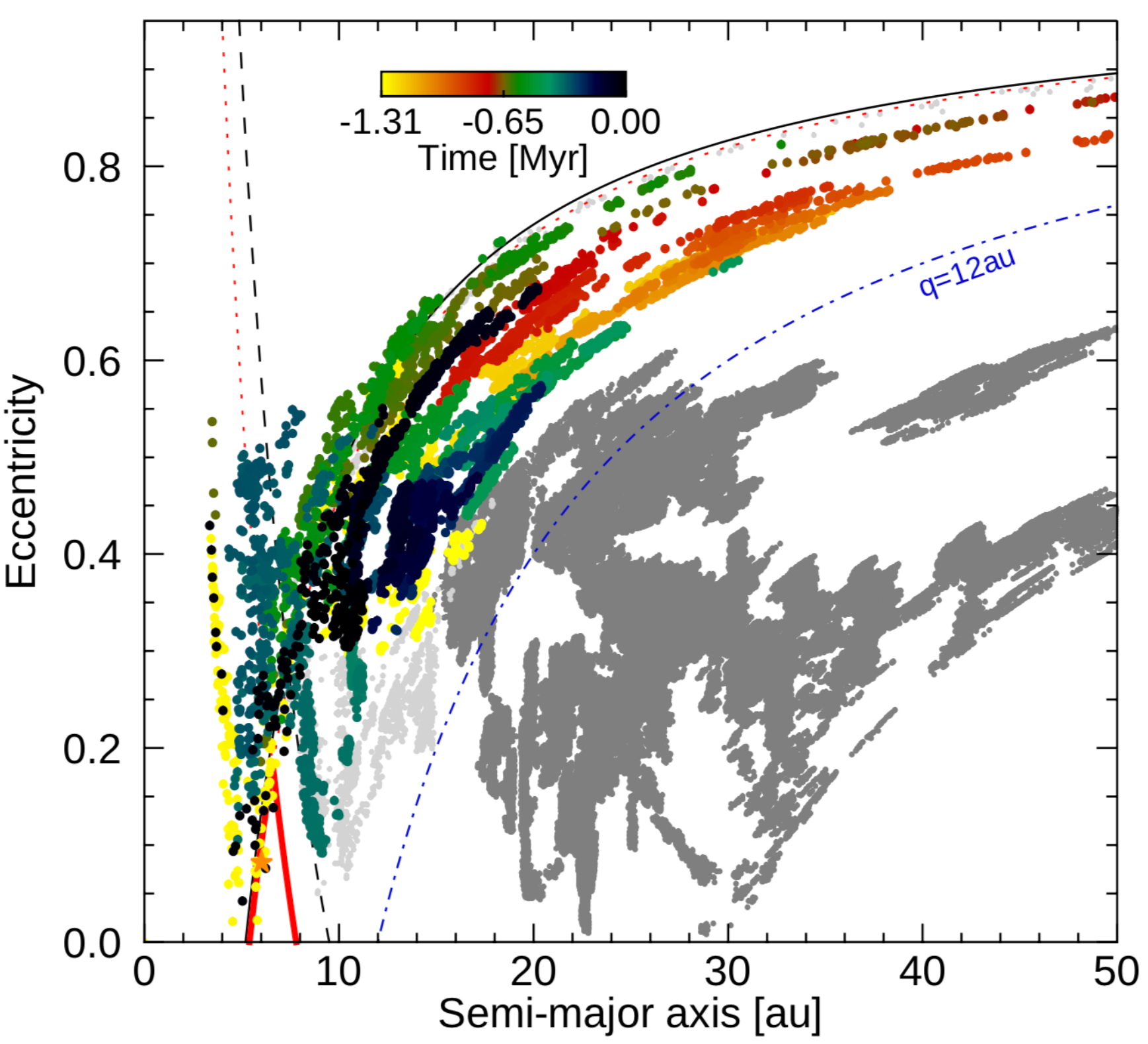}
   \caption{Orbital evolution of two clones of comet 29P in the eccentricity-semimajor axis plane.  The clone from the top panel underwent relatively little heating before approaching the orbit of comet 29P, whereas the clone from the bottom panel was strongly heated during close passages to the Sun. 
   One data point is given every 100~yr of dynamical evolution: the color code provides the time evolution for the duration subsequently displayed in Fig.\ref{qt29P}. The orbital evolution prior and after this time subset is shown in dark and light grey respectively. The black solid line shows orbits with a perihelion distance of 5.2au, the black dashed line shows orbits with an aphelion distance of 9.4au, and red lines provide the limits of the gateway. A blue dot-dashed line shows orbits with a perihelion distance of 12au.}
   \label{ae_29P}
\end{figure}

\begin{figure}[!h]
   \centering
   \includegraphics[width=\columnwidth]{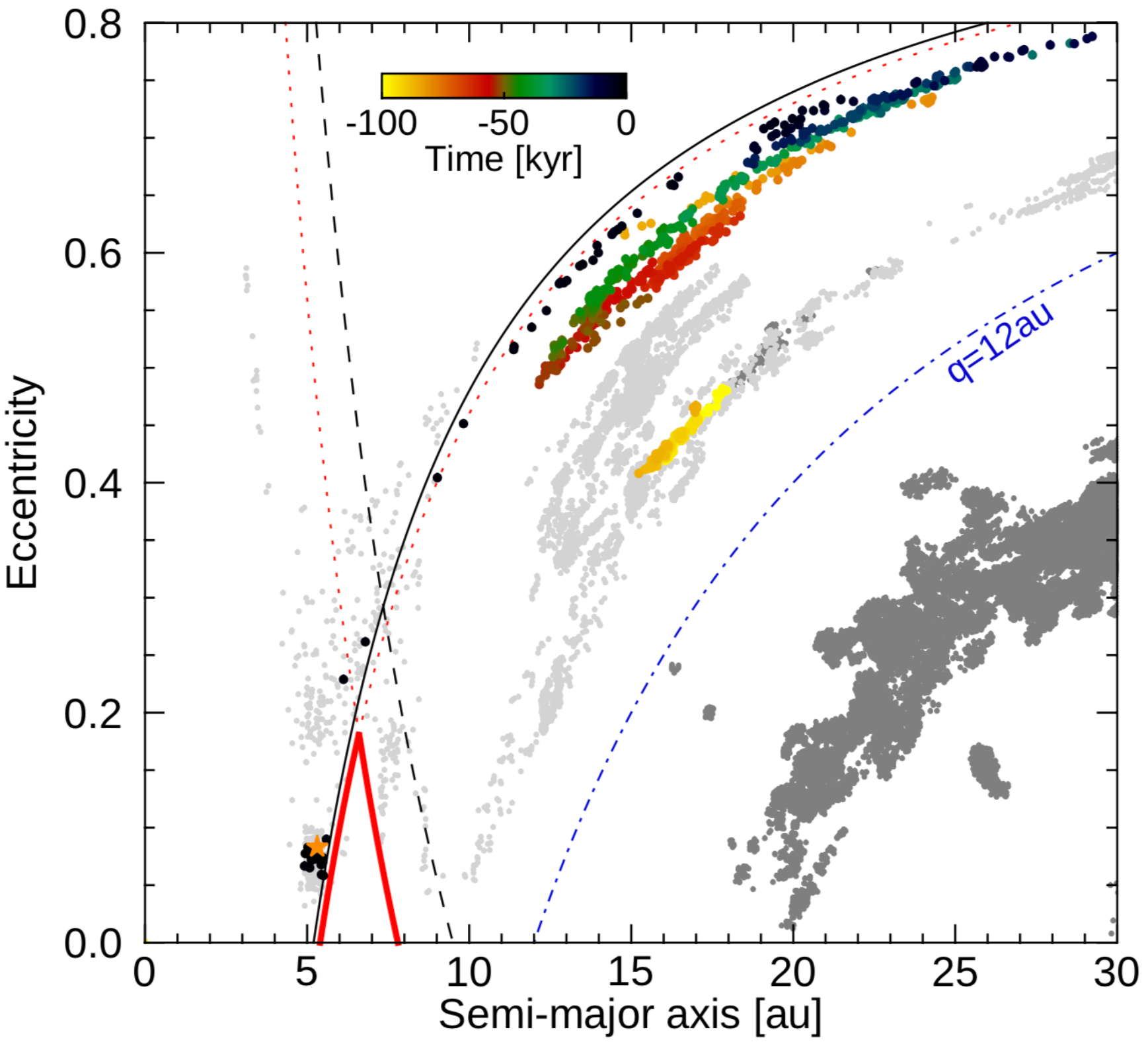}
   \includegraphics[width=\columnwidth]{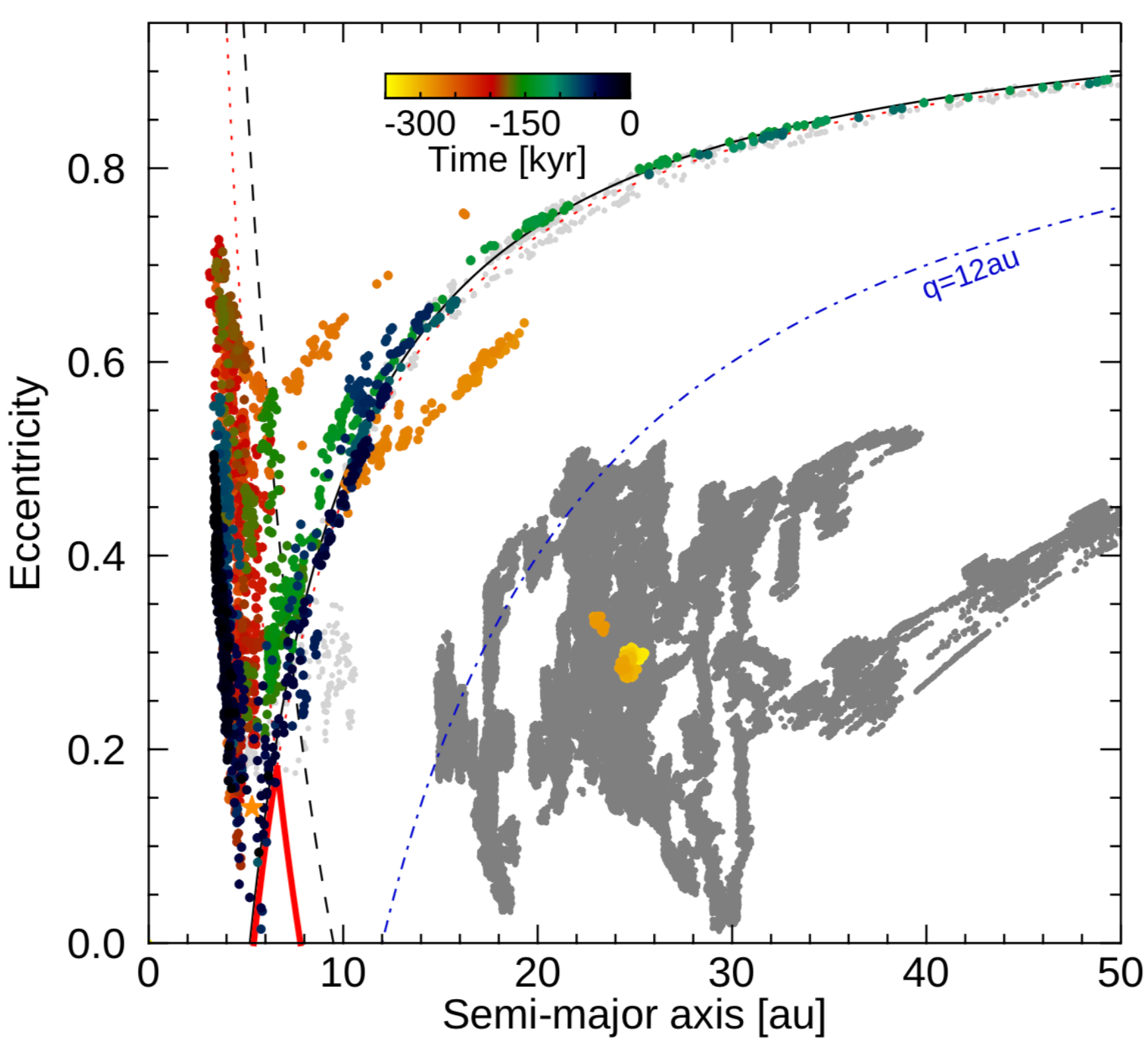}
   \caption{Orbital evolution of two clones of comet LD2 in the eccentricity-semimajor axis plane.  As in Fig.~\ref{ae_29P}, the clone from the top panel underwent relatively little heating before approaching the orbit of comet LD2, whereas the clone from the bottom panel was strongly heated during close passages to the Sun.
   One data point is given every 100~yr of dynamical evolution: the color code provides the time evolution for the duration subsequently displayed in Fig.\ref{qtLD2}. The orbital evolution prior and after this time subset is shown in dark and light grey respectively. The black solid line shows orbits with a perihelion distance of 5.2au, the black dashed line shows orbits with an aphelion distance of 9.4au, and red lines provide the limits of the gateway. A blue dot-dashed line shows orbits with a perihelion distance of 12au.}
   \label{ae_LD2}
\end{figure}

\begin{figure}[!h]
   \centering
   \includegraphics[width=\columnwidth]{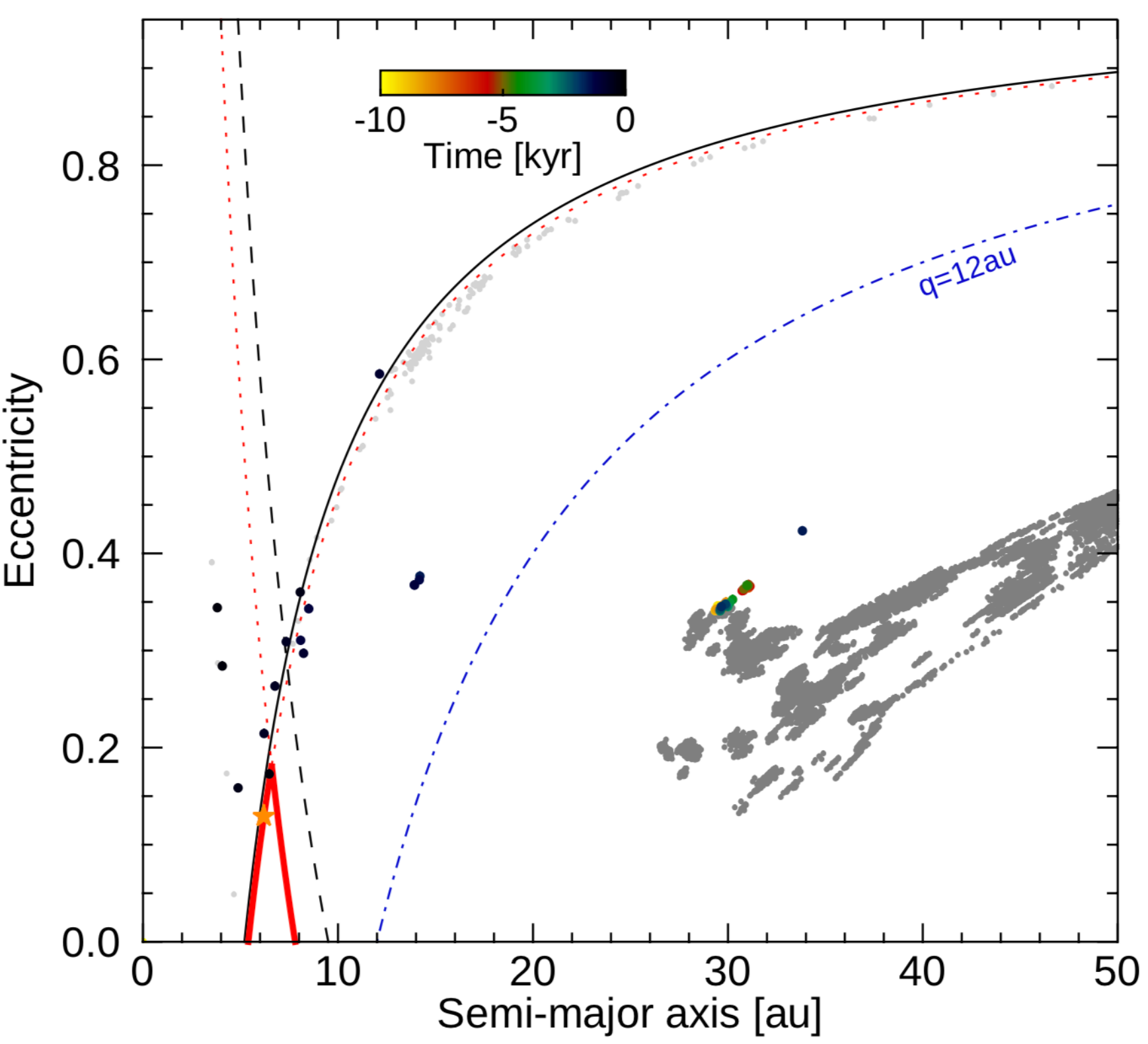}
   \includegraphics[width=\columnwidth]{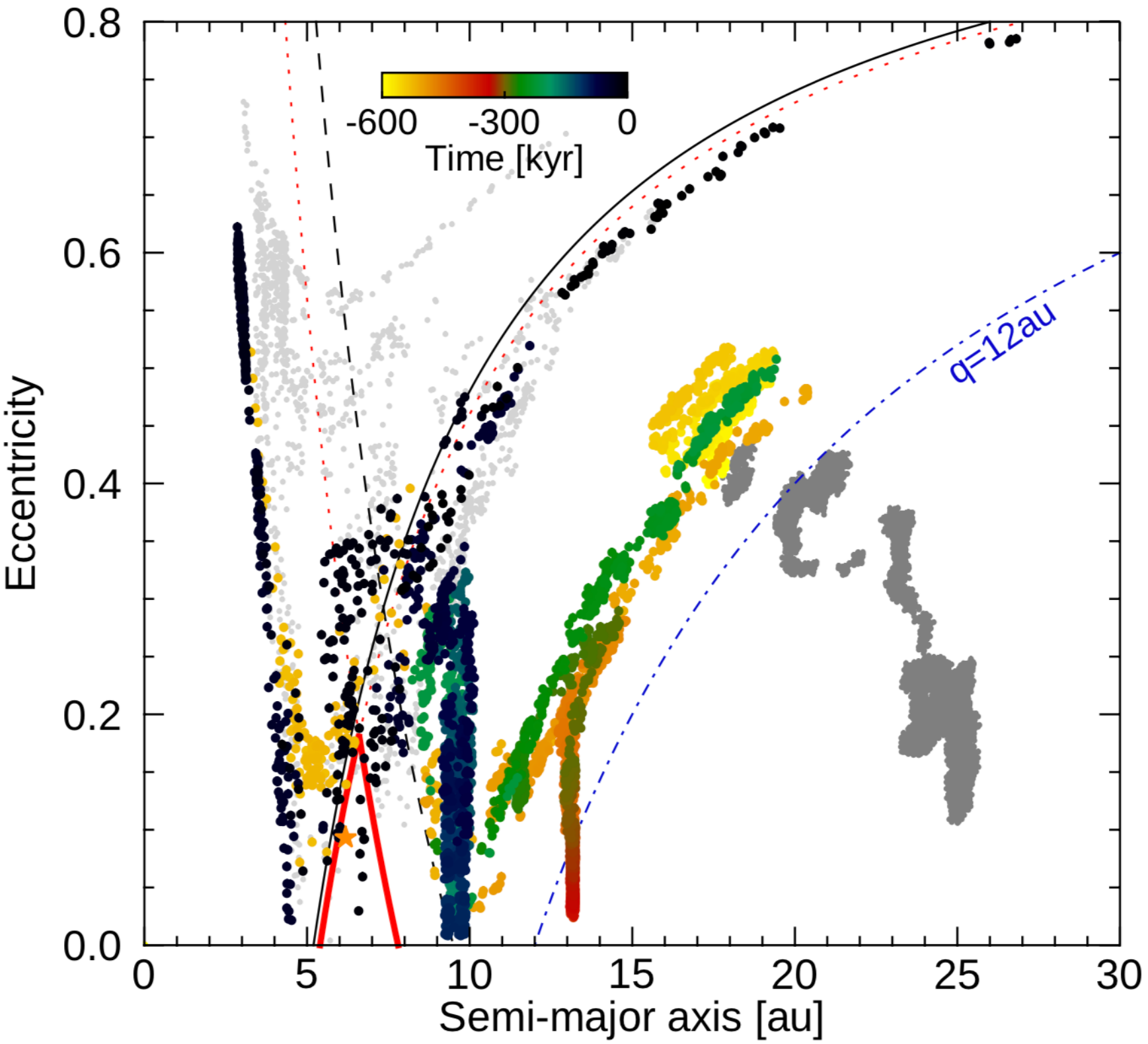}
   \caption{Orbital evolution of two clones of comet CL94 in the eccentricity-semimajor axis plane.  As in Fig.~\ref{ae_29P}, the clone from the top panel underwent relatively little heating before approaching the orbit of comet CL94, whereas the clone from the bottom panel was strongly heated during close passages to the Sun.
   One data point is given every 100~yr of dynamical evolution: the color code provides the time evolution for the duration subsequently displayed in Fig.\ref{qtCL94}. The orbital evolution prior and after this time subset is shown in dark and light grey respectively. The black solid line shows orbits with a perihelion distance of 5.2au, the black dashed line shows orbits with an aphelion distance of 9.4au, and red lines provide the limits of the gateway. A blue dot-dashed line shows orbits with a perihelion distance of 12au.}
   \label{ae_CL94}
\end{figure}

For our three objects, several features of interest are observed. First, the lifetime of clones spans a wide range of values. For 29P, some clones transition from 30~au to the box as fast as $\sim$4.5~Myr, while others take much longer than 100~Myr to reach 29P's orbit from the outer solar system (the longest being 375.5~Myr). For LD2, the dynamical timescales range from 16.6 to 480.5~Myr, and for CL94, the lifetime range is even larger: from 9.3 to 878.1~Myr. Figures \ref{ae_29P}, \ref{ae_LD2} and \ref{ae_CL94} provide details on the evolution of orbital elements for clones of 29P, LD2 and CL94 respectively, before they enter the designated box.
The dynamical behavior of each clone is unique and chaotic, with sometimes long periods spent in regions where phase transitions would be expected to occur \citep[see][and Fig. \ref{ae_29P} to \ref{ae_CL94}]{Guilbert2012, Davidsson2021}.
Second, we see that before entering the designated box, most clones explored orbits with lower perihelion distances (either as JFCs or Jupiter-crossers). Indeed, once their orbital evolution is dominated by gravitational interactions with Jupiter, changes in the perihelion distance become more frequent and chaotic. 
Third, some clones can enter the orbital box defined for each object several times during their lifetime.
Furthermore, we notice that clones can enter the orbital box of two different objects throughout their lifetime. For instance, we have one clone in common between 29P and LD2, three clones in common between 29P and CL94, and two clones in common between LD2 and CL94.

\subsection{Thermal evolution and internal structure}

\begin{figure}[!ht]
    \centering
   \includegraphics[width=\columnwidth]{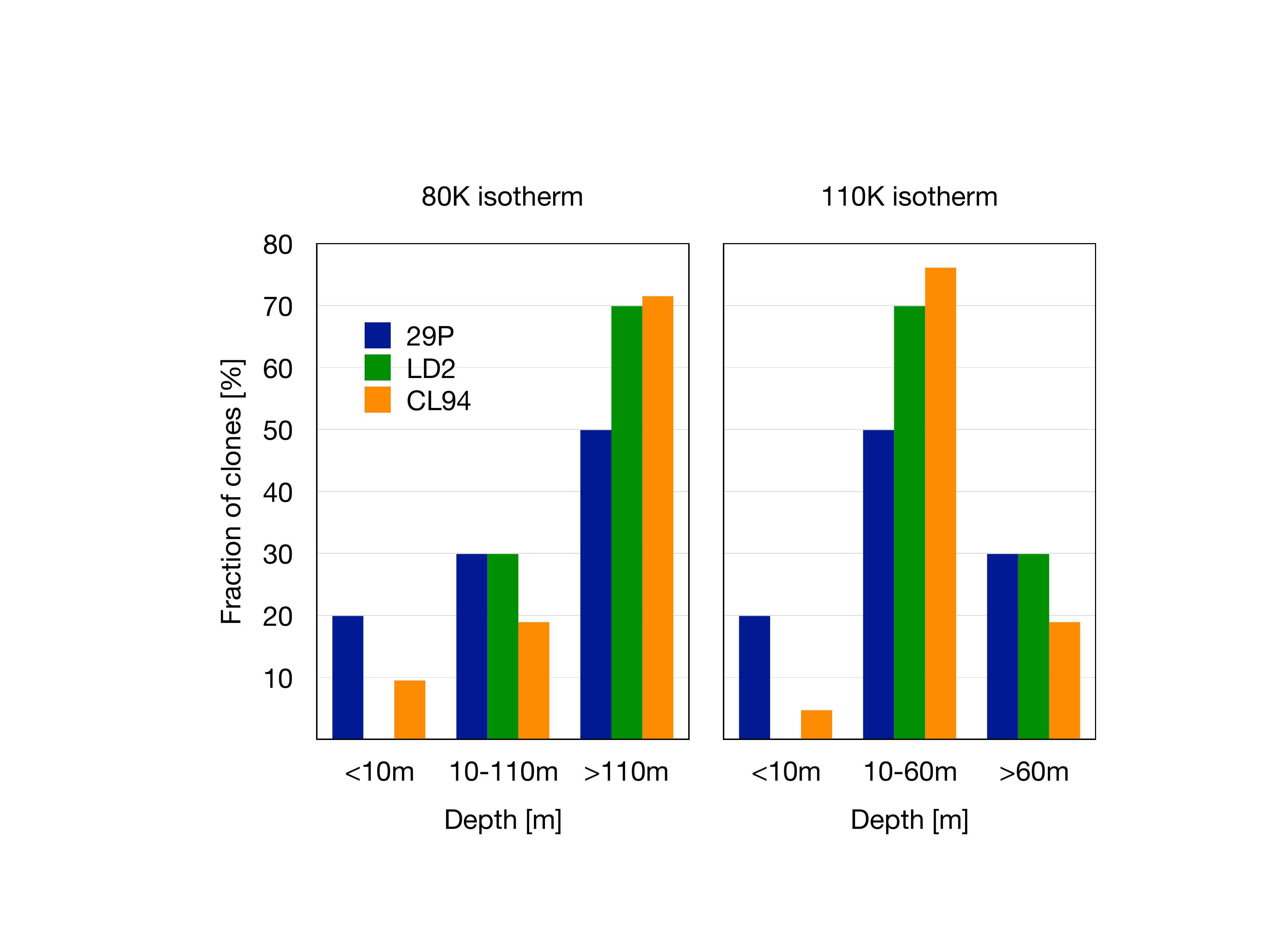}
   \caption{Temperature distributions for clones of  29P (blue), LD2 (green) and CL94 (orange).}
   \label{distrib_clones}
\end{figure}

\begin{figure*}[!ht]
   \centering
   \includegraphics[width=\textwidth]{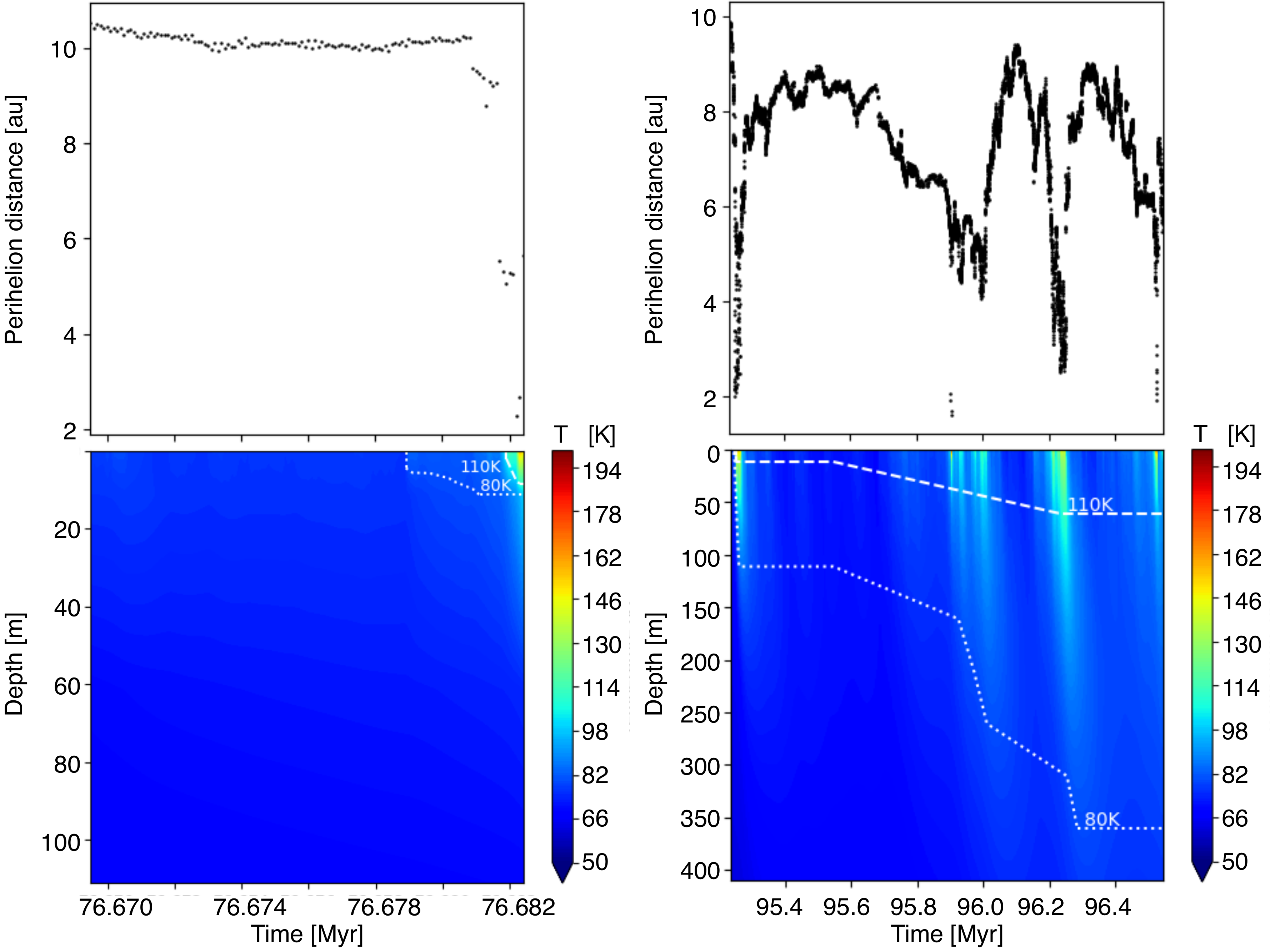}
   \caption{Distribution of the internal temperature as a function of depth and time for two clones of 29P, resulting from orbital evolution. Top panels show the evolution of perihelion distance as a function of time. Bottom panels show the resulting evolution of internal temperature. The least processed clone of 29P is shown on the left. The timeline focuses on the last 100~kyr of thermal and dynamical evolution, before the clone's orbital elements match those of 29P. On the right, one of the most processed clone is shown: the timeline focuses on the last 1~Myr of thermal and dynamical evolution, before the clone's orbital elements match those of 29P. The maximum depths of the 80~K and 110~K isotherms are shown with a white dotted and dashed line respectively.}
   \label{qt29P}
\end{figure*}

\begin{figure*}[!ht]
   \centering
   \includegraphics[width=\textwidth]{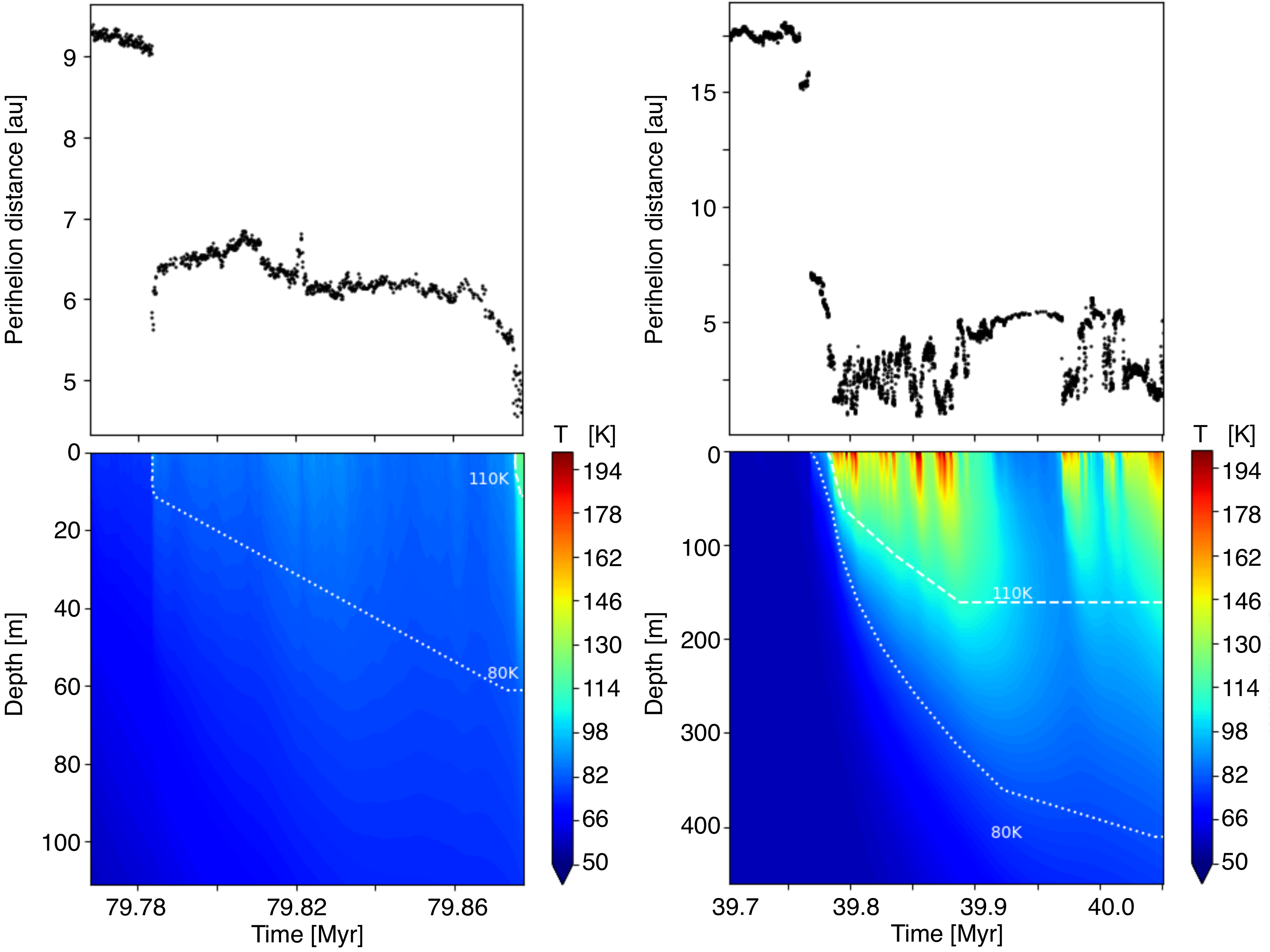}
   \caption{Distribution of the internal temperature as a function of depth and time for two clones of LD2, resulting from orbital evolution. Top panels show the evolution of perihelion distance as a function of time. Bottom panels show the resulting evolution of internal temperature. The least processed clone of LD2 is shown on the left. The timeline focuses on the last 100~kyr of thermal and dynamical evolution, before the clone's orbital elements match those of LD2. On the right, one of the most processed clone is shown: the timeline focuses on the last 300~kyr of thermal and dynamical evolution, before the clone's orbital elements match those of LD2. The maximum depths of the 80~K and 110~K isotherms are shown with a white dotted and dashed line respectively.}
   \label{qtLD2}
\end{figure*}

\begin{figure*}[!ht]
   \centering
   \includegraphics[width=\textwidth]{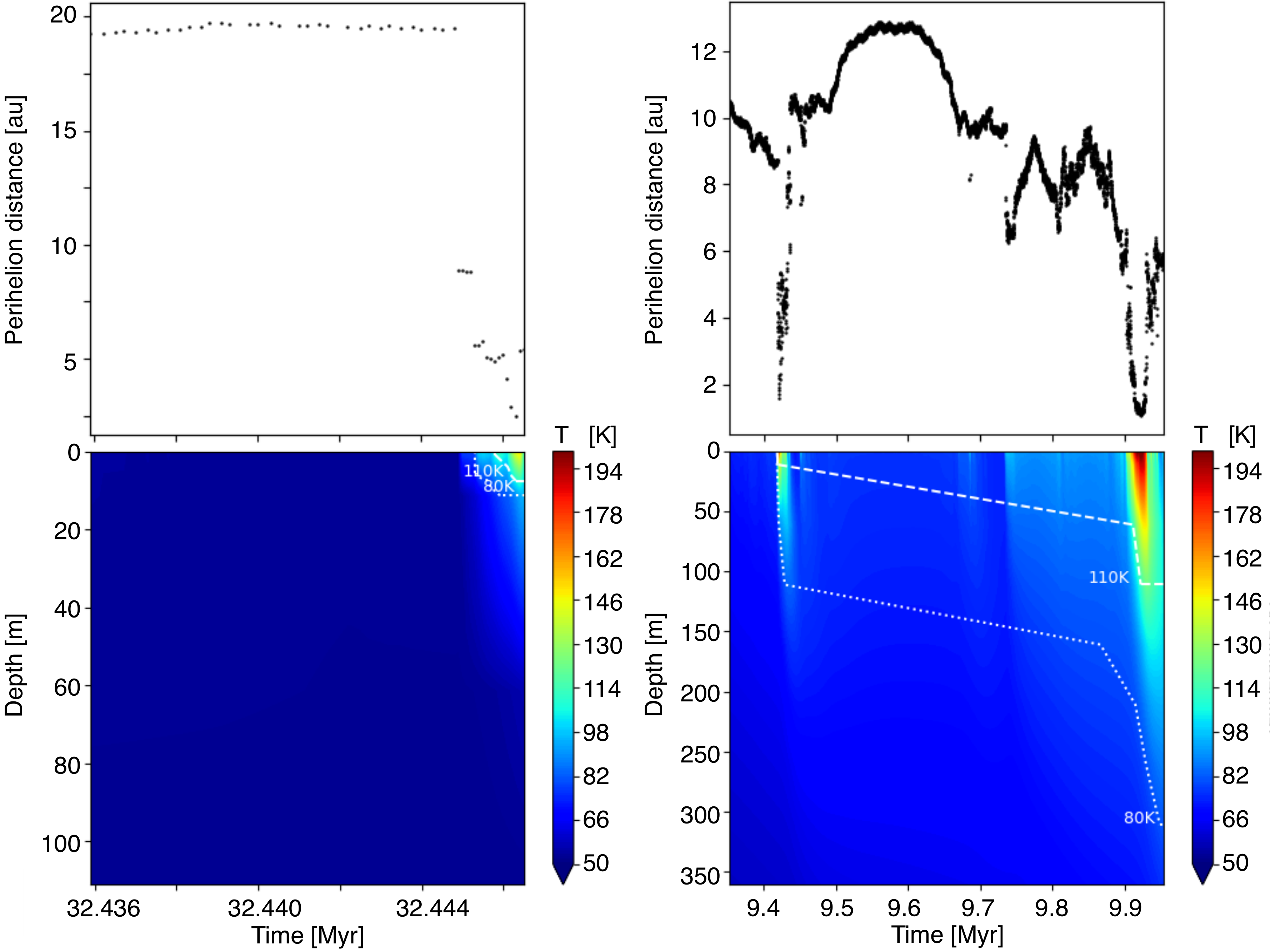}
   \caption{Distribution of the internal temperature as a function of depth and time for two clones of CL94, resulting from orbital evolution. Top panels show the evolution of perihelion distance as a function of time. Bottom panels show the resulting evolution of internal temperature. The least processed clone of CL94 is shown on the left, with a lifetime of $\sim$32.4~Myr. The timeline focuses on the last 10~kyr of thermal and dynamical evolution, before the clone's orbital elements match those of CL94. On the right, the most processed clone is shown, yet with a short lifetime ($\sim$10~Myr). The timeline focuses on the last 500~kyr of thermal and dynamical evolution, before the clone's orbital elements match those of CL94. The maximum depths of the 80~K and 110~K isotherms are shown with a white dotted and dashed line respectively.}
   \label{qtCL94}
\end{figure*}

Clones of individual objects experience a variety of timescales of residency in the giant-planet region, and of variations in orbital elements. This inevitably entails a diversity in thermal processing. The relationship between timescale and thermal processing is not straightforward though, and is clearly dependent upon the unique orbital track followed by each clone.  
In order to assess the degree of processing of each individual clone, we record the depth of the two isotherms of interest: their distributions in depth for clones of our three objects is shown in Fig.\ref{distrib_clones}.
Visualizing how the heat propagates below the surface is also informative, so we show in Figures \ref{qt29P}, \ref{qtLD2} and \ref{qtCL94} two examples of thermal evolution coupled to the dynamical evolution for clones of 29P, LD2 and CL94 respectively. These correspond to clones whose orbital evolutions were presented in Fig.\ref{ae_29P} to \ref{ae_CL94}. For each Centaur of interest, we have selected one clone arriving relatively unaltered in the designated box, and one comparatively thermally-processed clone.

Centaur 29P is the object, amongst our three, with the largest fraction of relatively unaltered clones: two of them, despite having quite long lifetimes ($\sim$29 and 76.7~Myr respectively), reach the 29P current position with an unaltered composition. Their interiors are barely heated above 80~K in the upper 10~m. Three clones have the 80~K and 110~K isotherms located around 50~m and 10~m deep respectively, while half of the 29P clones have been processed to the extent that the 110~K isotherm is located below 50~m, and the 80~K isotherm below 100~m or more (up to $\sim$360~m). 
For clones of LD2, the depths for the 80~K and 110~K isotherms are statistically greater than for 29P, since no LD2 clone reaches the orbital box as unprocessed as the two 29P clones mentioned above. 
Of the 10 clones of LD2, all are thus moderately to substantially processed, with the 80~K and 110~K isotherms found at least below 10~m, and mostly below 100~m for the 80~K isotherm. For the two most processed clones, with two different lifetimes of more than 480~Myr vs. 40~Myr, the 80~K isotherm is located beyond 400~m when they arrive in the LD2 orbital box.
Of the CL94 clones, only one remains relatively unprocessed (with the 80 and 110~K isotherms within the uppermost 10~m), while for most the 80~K and 110~K isotherms are located beyond 110~m and 60~m respectively.


\section{Discussion}

\subsection{On transitioning through the gateway}

By means of forward modeling of the dynamical cascade from TransNeptunian Objects to JFCs, \citet{Sarid2019} suggested that a specific dynamical pathway should facilitate the transition between the Centaur and JFC populations. 
They found that 21\% of Centaurs transition to JFC orbits through the gateway (30\% when adding the gravitational perturbations from inner solar system planets). Because of stochastic gravitational perturbations from Jupiter, objects can jump in and out of the giant planet region several times in their lifetime, so that more than 3/4 of them would eventually go through the gateway, or nearly half of objects with a perihelion distance smaller than 3au.

We use a sample of simulated active and visible JFCs, all with perihelion distances smaller than 2.5~au at some point in their lifetime. 
Our results can be directly compared with those of \citet{Sarid2019}, as they provide for statistics for clones reaching $q<$~3~au. They find that nearly half of those clones spend some time in the gateway, which is consistent with the 54.6\% of clones we find. In terms of pure dynamical pathways, our results are thus completely aligned with those of \citet{Sarid2019}. 

However, when it comes to the pristine nature of these objects, the thermal processing sustained prior to their passage in the gateway, hence the direction from which they enter the gateway, matters. When we constrain the origin of clones the first time they reach the gateway, we find that strictly speaking, our population has only 20.9\% of Centaurs which actually go through the gateway prior to becoming JFCs for the first time, again consistent with  \citet{Sarid2019}.
Since 159 clones (45.4\%) never go through the gateway at any point of their lifetime, we find that most Centaurs (79.2\%) transition \textbf{from} the giant planet region \textbf{to} the JFC population outside of the gateway region.

As a result, objects in the gateway are statistically more prone to being thermally processed than other Centaurs crossing the orbit of Jupiter for the first time, because a higher fraction of gateway objects have already been processed as JFCs. \citet{Guilbert2016} suggested that typical JFCs could have their subsurface altered down to a few hundred meters before entering the inner solar system. \citet{Gkotsinas2022} found that due to the stochastic nature of comet trajectories toward the inner solar system, JFCs can experience multiple heating events resulting in substantial chemical alteration of the upper layers, down to several hundred meters. 
Accessing material below this depth would require the cumulative erosion effect of multiple perihelion passages as JFCs \citep[typically a few meters per perihelion passage, ][]{Prialnik2004, Huebner2006}. Therefore, for most JFCs, outgassing observed today might occur from a layer thermally altered during the Centaur stage. Because several transitions between the Centaur and the JFC populations are possible \citep[][and this work]{Sarid2019, Gkotsinas2022}, the cumulative effect of multiple transitions should lead to a complex internal structure and composition. Our results suggest that statistically, there is a $\sim$50\% chance that any object in the gateway is one of these processed bodies.

\subsection{On objects currently in the gateway}

To illustrate the latter point, we focused on three objects of interest: Centaur 29P/Schwassmann-Wachmann~1 and comets P/2019~LD2~(ATLAS), 423P/Lemmon (2008 CL94) currently residing in the gateway. For the first two, the dynamical evolution of many dynamical clones was integrated backwards in time \citep{Sarid2019, Steckloff2020}. This method can only inform on a recent past, typically hundreds to thousands of years, before stochastic gravitational interactions with Jupiter make clones diverge. These backward integrations suggest that it is unlikely they should have spent any significant amount of time in the inner solar system. As a result, they should not have experienced any significant thermophysical evolution, and their current activity would be representative of nearly-pristine objects.

Our results stem from a different methodology, arguably the only reliable strategy, i.e. forward modeling of the dynamical evolution \citep{Morbidelli2020}. Our statistics are necessarily limited, given the number of clones for 29P and LD2 in particular. However, there are issues in using backward integrations to investigate the evolution of solar system objects, as detailed by \citet{Morbidelli2020}. Therefore, even if JFC clones from \citet{Nesvorny2017} yield a smaller number of clones of each individual object of interest compared to the aforementioned studies, the overall sample has strengths that cannot be excluded. For instance, our results span a much longer period, since our dynamical tracks follow objects from the time they leave the transneptunian region for several million to hundred million years. Our analysis justifies that a broad look at the time spent in the giant planet region is not sufficient for assessing the thermal evolution of Centaurs. Instead, the detailed orbital evolution, and the resulting thermal processing, must be constrained for each object. We find that each object of interest has a higher than 50\% chance that the layer currently contributing to its observed activity has been physically and chemically altered, due to thermal processing sustained during previous stages of evolution.

\subsection{Significance for the Centaur and JFC populations}
Understanding the mechanisms at the origin of Centaurs' activity is paramount to fully comprehend the extent of the post-formation thermal processing of JFCs.
Current impediments for fulfilling that goal come, on one hand, from the lack of volatile detection in their coma. 
No strong detection of gaseous CO has been made to date \citep[e.g.][]{Drahus2017}, except for 29P \citep{Senay1994, Crovisier1995, Gunnarsson2008, Wierzchos2020}. CO was marginally detected in the coma of 2060~Chiron at 8.5~au \citep{Womack1999, Womack2017}, as well as on the coma of 174P/Echeclus during an outburst at 6.1 au \citep{Wierzchos2017}. 
These observations are consistent with the activity of Centaurs not being driven by the sublimation of CO, but such limited dataset does not allow to constrain phase transitions at the source of outgassing from nuclei.  
This may be due to current observational sensitivities, which will improve in the JWST and ALMA era. 
On the other hand, \citet{Cabral2019} argued that we currently do not have an appropriate dataset to constrain the origin of Centaurs’ activity, because no survey has ever been dedicated to this population. 
Compared to TransNeptunian Objects, Centaurs have a different detectability in motion-rate dependent surveys. For instance, a survey such as OSSOS \citep[Outer Solar System Origins Survey, ][]{Bannister2016} has an observation cadence biased toward detecting dynamically stable orbits beyond Saturn \citep{Tiscareno2003, diSisto2010}.
As of today, no survey has adequately targeted the motion rate of objects in the 5-12~au region, where Centaurs with more unstable orbits can be found, and active Centaurs are currently observed, in a well-characterized manner. 

As for any Centaur or JFC, the activity of 29P, LD2 and CL94 reflects the composition and structure inherited from these previous stages of evolution. As such, our results suggest that current outgassing likely arises from a layer significantly altered prior to observations. Depending on thermophysical parameters, compositions, and other poorly constrained properties, it is still possible that the effects of such thermal processing could be limited to a modest near-surface layer, for 29P in particular. 
Because the degree of activity, reflected for example by production rates, may not be straightforwardly linked to the degree of processing experienced by comets prior to current observations, it is important to remain cautious when claiming that any active Centaur is representative of a pristine nucleus. Our results suggest that objects in the gateway may not have a particular significance, compared to other active Centaurs, for a better understanding of this population or the onset and development of activity in the giant-planet region. In any case, some caution ought to be applied when claiming that gateway Centaurs and their activity are representative of the onset of activity experienced by supposedly pristine objects, prior to their transition on JFC orbits

\section{Summary}

We aim to constrain the internal thermal structure resulting from the orbital evolution of Centaurs, prior to their transition in the JFC region, whether this transition occurs through the gateway or not.   
We used simulation outcomes of the coupled thermal and dynamical evolution from \citet{Gkotsinas2022}, for a population of JFC clones from \citet{Nesvorny2017}. We find that:
\begin{itemize}
    \item[1)] Only $\sim$20\% Centaurs go through the gateway prior to transitioning on JFC orbits. Most Centaurs in our sample make their first transition to the JFC population from outside of this region.
    \item[2)] More than half of the dynamical clones entering the gateway for the first time have already been JFCs. Statistically, objects in the gateway are thus more processed than the rest of the objects when they start transitioning on JFC orbits.
    \item[3)] 29P/Schwassmann-Wachmann1,  P/2019 LD2 (ATLAS), and P/2008 CL94 (Lemmon) have a higher than 50\% chance to be thermally processed. As a result, the layer currently contributing to their observed activity could be physically and chemically altered, and not representative of the initial state of these objects.
\end{itemize}

\begin{acknowledgements}

We thank Darryl Seligman for his excellent review of our work, helping us to improve this manuscript. We warmly thank members of the ISSI (International Space Science Institute) team led by Rosita Kokotanekova, for constructive discussions on the Centaur population. This study is part of a project that has received funding from the European Research Council (ERC) under the European Union’s Horizon 2020 research and innovation programme (Grant agreement No. 802699). We gratefully acknowledge support from the PSMN (Pôle Scientifique de Modélisation Numérique) of the ENS de Lyon for computing resources.  

\end{acknowledgements}

\facility{PSMN, ENS de Lyon}

\appendix

\begin{figure*}[!ht]
   \centering
   \includegraphics[width=\textwidth]{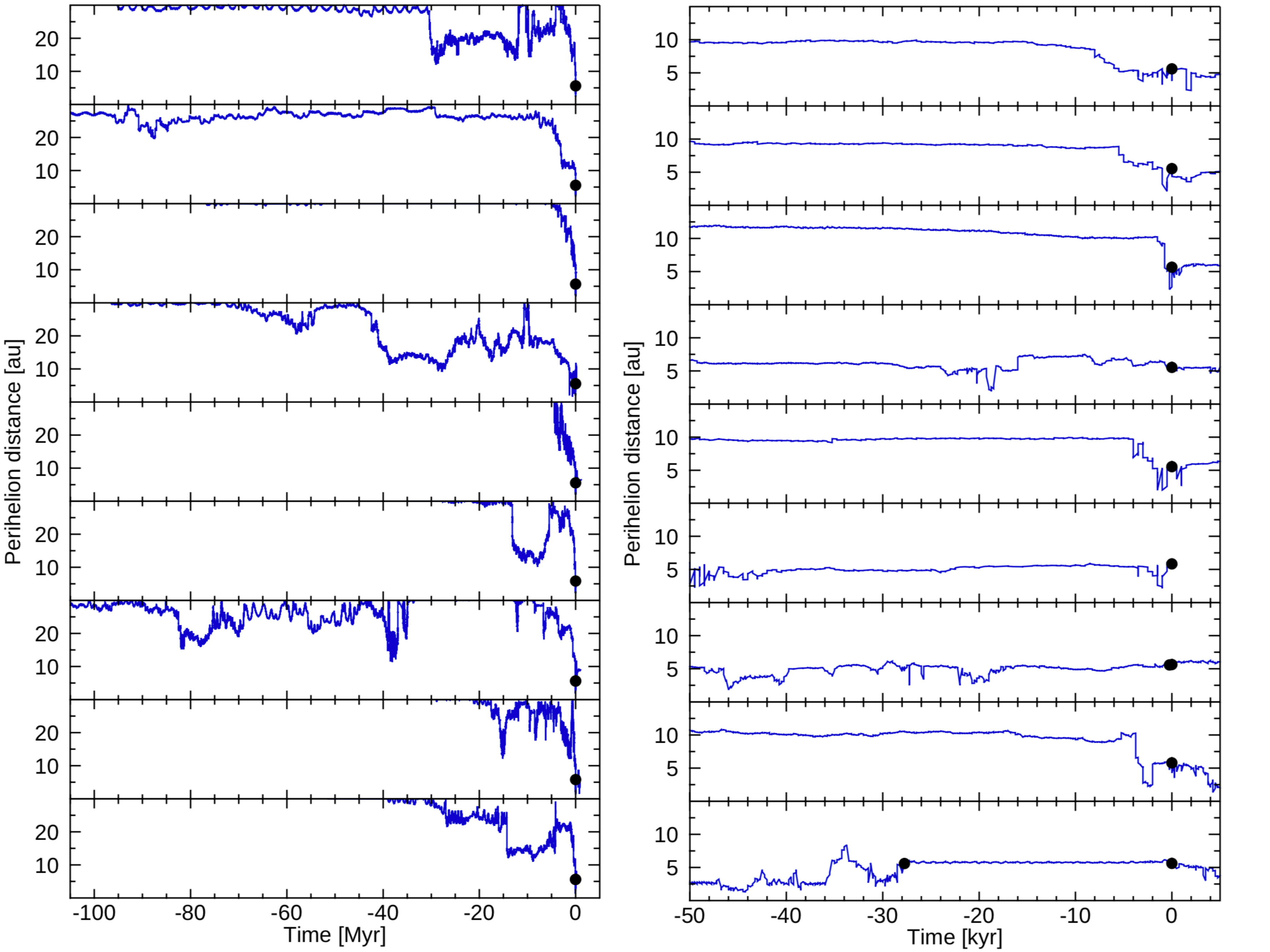}
   \caption{Evolution of the perihelion distance for 9 clones of 29P, within the last 100~Myr (left) and 50~kyr (right) of their orbital evolution toward ``becoming'' 29P -- marked as black dots. Reference time 0 is chosen as the last time the clone enters the ``29P box'' of accepted orbital elements.}
   \label{29P_dyn}
\end{figure*}

\begin{figure*}[!h]
   \centering
   \includegraphics[width=\textwidth]{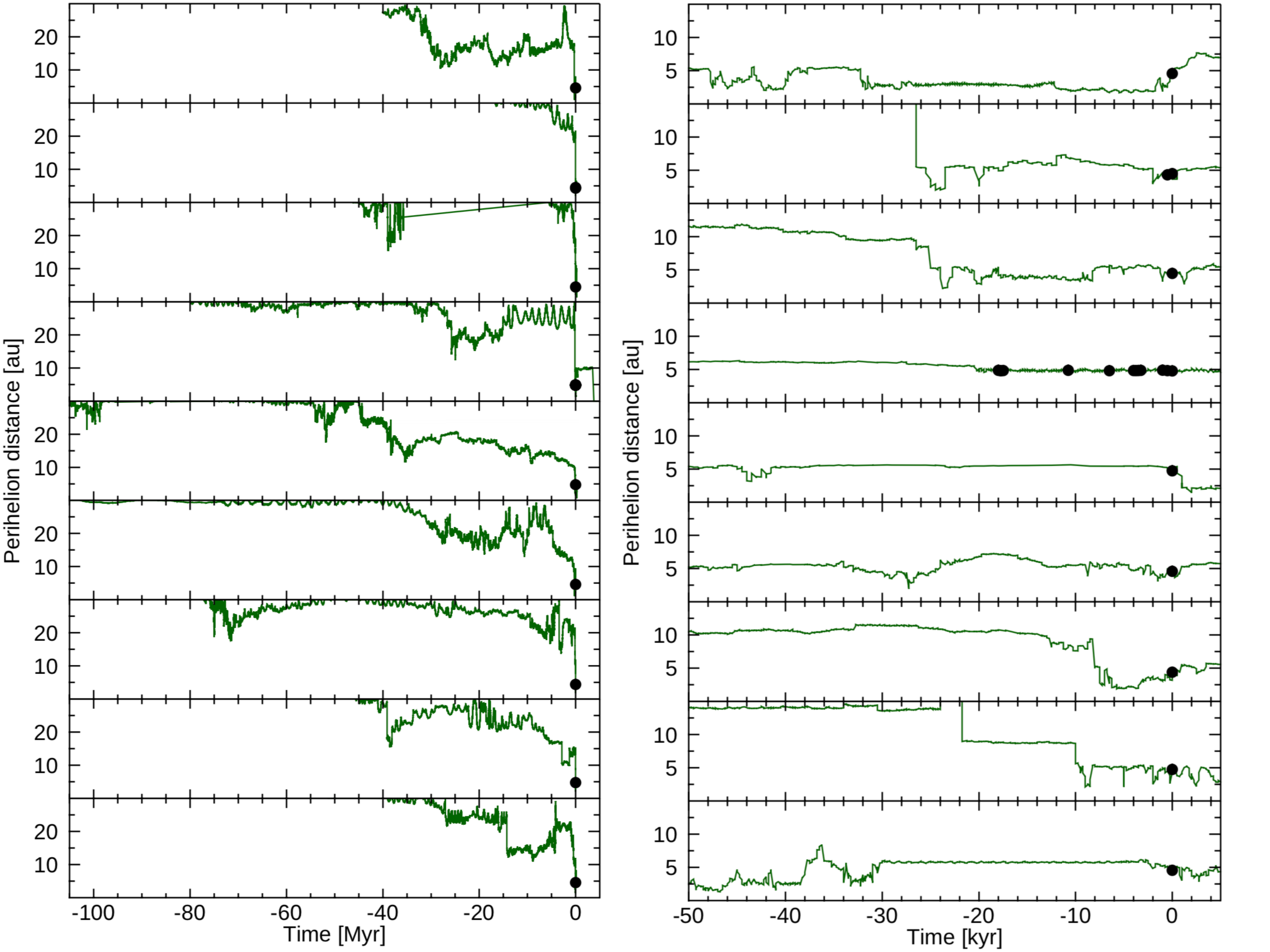}
   \caption{Evolution of the perihelion distance for 9 clones of LD2, within the last 100~Myr (left) and 50~kyr (right) of their orbital evolution toward ``becoming'' LD2 -- marked as black dots. Reference time 0 is chosen as the last time the clone enters the ``LD2 box'' of accepted orbital elements.}
   \label{LD2_dyn}
\end{figure*}

\begin{figure*}[!h]
   \centering
   \includegraphics[width=\textwidth]{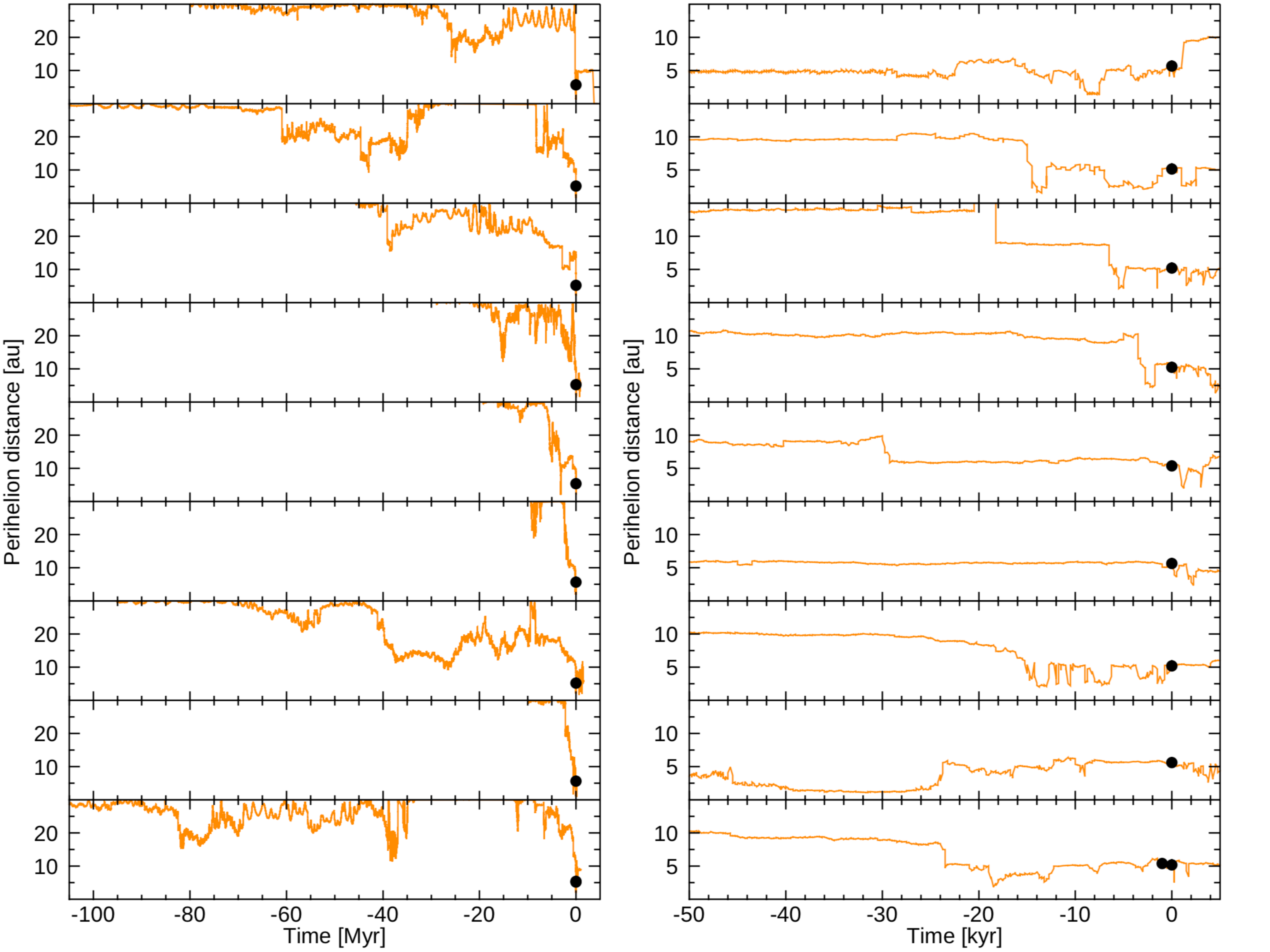}
   \caption{Evolution of the perihelion distance for 9 clones of CL94, within the last 100~Myr (left) and 50~kyr (right) of their orbital evolution toward ``becoming'' CL94 -- marked as black dots. Reference time 0 is chosen as the last time the clone enters the ``CL94 box'' of accepted orbital elements.}
   \label{CL94_dyn}
\end{figure*}

To illustrate the diversity of dynamical behaviors, we show in Figures \ref{29P_dyn}, \ref{LD2_dyn} and \ref{CL94_dyn} the evolution of the perihelion distance for 9 clones of 29P, LD2 and CL94 respectively. 

\bibliography{gateway_accepted}{}
\bibliographystyle{aasjournal}

\end{document}